\documentclass[aps,pre,twocolumn,longbibliography]{revtex4-1}
\usepackage{mathtools}
\usepackage{amsfonts,amssymb,wasysym} 
\setcitestyle{super,comma,sort&compress}
\makeatletter
\def\blfootnote{\xdef\@thefnmark{}\@footnotetext}
\makeatother
\usepackage[english]{babel}
\usepackage[utf8x]{inputenc}
\usepackage{graphicx}
\usepackage[colorinlistoftodos]{todonotes}
\begin{document}
\title{Connecting the quantum and classical worlds}
\author{Barbara Drossel}
\affiliation{Institute of Condensed Matter Physics, Darmstadt University of Technology}

\begin{abstract}
By considering (non-relativistic) quantum mechanics as it is done in practice in particular in condensed-matter physics, it is argued that a deterministic, unitary time evolution within a chosen Hilbert space always has a limited scope, leaving a lot of room for embedding the quantum-classical transition into our current theories without recurring to difficult-to-accept interpretations of quantum mechanics. Nonunitary projections to initial and final states, the breaking of time-reversal symmetry, a change of Hilbert space, and the introduction of classical concepts such as external potentials or localized atomic nuclei are widespread in quantum mechanical calculations. Furthermore, quantum systems require classical environments that enable the symmetry breaking that is necessary for creating the atomic configurations of molecules and crystals. This paper argues that such classical environments are provided by finite-temperature macroscopic systems in which the range of quantum correlations and entanglement is limited. This leads to classical behavior on larger scales, and to  collapse-like events in all dynamical processes that become coupled to the thermalized degrees of freedom.
\end{abstract}
\maketitle
\section{Introduction}
The quantum and the classical world are fundamentally different. 
Nonrelativistic quantum mechanics is governed by the Hamilton operator of a many-particle system,
\begin{equation}
\hat H= \sum_{\alpha=1}^N \frac{\hat p_\alpha^2}{2m_\alpha} + \frac 1 2\sum_{\alpha\neq\beta}V(\vec r_\alpha - \vec r_\beta) \, ,\label{fullhamiltonian}
\end{equation}
where  $\alpha$ and $\beta$ count the particles of the system (for instance atoms, or nuclei and electrons) and where the potential $V(\vec r_\alpha - \vec r_\beta)$ describes the pairwise interactions between the particles. 
 The associated Schr\"odinger equation 
\begin{equation}
i \hbar \frac{\partial}{\partial t}\psi(\vec r_1, \cdots, \vec r_N,t) = \hat H \psi(\vec r_1, \cdots, \vec r_N,t) \label{fullschroedinger}
\end{equation}
fully determines the time evolution of a system. Given an intial state $\psi(\vec r_1, \cdots, \vec r_N,0)$, the future (and also the past) states are given by
\begin{equation}
\psi(\vec r_1, \cdots, \vec r_N,t) = e^{-i\hat H t/\hbar}\, \psi(\vec r_1, \cdots, \vec r_N,0). \label{timeevolution}
\end{equation}
Such a unitary time evolution implies an entanglement of all particles that are coupled via interaction potentials $V(\vec r_\alpha-\vec r_\beta)$: Using a one-particle basis $\{\phi_{i_\alpha}(\vec r_\alpha)\}$, one can expand the wave function $\psi(\vec r_1, \cdots, \vec r_N,t) $  in terms of products of one-particle wave functions, \begin{equation}\psi(\vec r_1, \cdots, \vec r_N,t)= \sum_{\{i_1,\dots,i_N\}} c_{i_1,\dots,i_N}(t)\prod_{\alpha=1}^N \phi_{i_{\alpha}}(\vec r_\alpha)\,  , \label{expansion}\end{equation}
where now the complex coefficients $c_{i_1,\dots,i_N}$ specify the state of the system.  If a particle was not entangled with the other particles, its state $\sum_{i_\alpha}c_{i_\alpha}(t)\phi_{i_\alpha}(\vec r_\alpha)$ would be a global factor in Eq.~(\ref{expansion}). However, such a product state satisfies the Schr\"odinger equation (\ref{fullschroedinger}) only if the particle does not interact with the other particles. Similarly, Eq.~(\ref{expansion}) can be written as a product of groups of particles only if there are no interactions between these groups.


In contrast, the macroscopic world surrounding us bears no resemblance to the unitary, entangled world suggested by quantum mechanics. Instead, it contains well localized objects that show no evidence of entanglement with other objects, is not invariant under time reversal, not linear, and not deterministic.

There are essentially two approaches to this contradiction: The first consists in accepting the validity of the mathematical description given above even for macroscopic systems. 
The Hamiltonian (\ref{fullhamiltonian}), together with the Schr\"odinger equation (\ref{fullschroedinger}), is then the "Theory of everything" of nonrelativistic condensed matter physics according to a well-known paper by Laughlin and Pines\cite{laughlin2000theory}. This means that a unitary time evolution in a huge Hilbert space is believed to somehow describe all possibilities of how nuclei and electrons can arrange themselves to form larger objects or move around in space. In particular, efforts are made to derive statistical mechanics from many-particle quantum mechanics \cite{Emch2007-EMCQSP, lindenpopescuetal2010,ShortFarrelly2012,reimann2013quantum,eisert2015quantum,gogolin2016equilibration}. A central role in these calculations is played by the concept of decoherence, which states that entanglement of a system with its environment destroys quantum superpositions within the system. Critics of this first approach point out that the calculations always include assumptions such as statistical independence and "typicalness" that are foreign to a deterministic theory\cite{kastner2014einselection,drossel2015relation}, and that the calculations alone cannot describe the objective occurrence of definite, random outcomes of individual measurements\cite{adler2003decoherence,schlosshauer2005decoherence,ellis2014evolving}.
Consequently, these calculations  are combined with interpretations of quantum mechanics such as many worlds\cite{everett1957relative},  relational\cite{rovelli1996relational}, consistent (or decoherent) histories\cite{griffiths1984consistent}, or statistical interpretations\cite{Ballentine}. These interpretations differ in the ontological status that they ascribe to the wave function and in their explanation of the observed randomness. 

The second approach to the contradiction between the quantum and classical world consists in accepting limits of validity to the description of many-particle systems by the Schr\"odinger equation \cite{petrosky1993poincare,leggett1992nature,ellis2012limits,bassi2013,tenreasons,gisin2016}. In fact, as emphasized by T. Leggett\cite{leggett1992nature} and R. Laughlin\cite{laughlin2000theory}, quantum mechanical calculations in condensed matter physics are never done using the full Hamiltonian, but effective theories. Furthermore, they depend on classical concepts and classical environments for those degrees of freedom that are not explicitely modeled \cite{ellis2012limits,chibbaro2014reductionism}. The view that there are limits to a unitary time evolution leads to interpretations such as the traditional Copenhagen interpretation and various collapse theories\cite{penrose1996gravity,bassi2013}. 

It is the purpose of this paper to push this second approach further by analyzing the meaning and limits of quantum mechanical calculations and emphasizing the important role of temperature at achieving the transition to classical behavior. In particular, a unitary time evolution requires the specification of the degrees of freedom of the system, which gives then a well-defined Hilbert space. However, the relevant degrees of freedom change when objects are formed, atoms become bound to form molecules or crystals, or particles are emitted from a system. The quantum mechanical description of such processes involves correspondingly a change of the Hilbert space.  While the Hilbert space of confined or bound objects has discrete energy levels, the mentioned processes involve the coupling to external, continuous degrees of freedom, and in particular to photons. We will see that whenever the coupling to a continuum is taken into account in quantum mechanical calculations, unitarity and time reversal symmetry are factually broken, as retarded propagators are chosen, localized sources are assumed to lie in the past but not in the future, and additional degrees of freedom are added and removed in a time-dependent manner. This becomes particularly relevant in macroscopic, finite-temperature systems, and we will argue that these systems must possess a finite length scale beyond which quantum entanglement is destroyed and interactions can be expressed in terms of classical potentials. This leads to a natural explanation of the boundary between quantum and classical mechanics and of the measurement process. 

This paper takes up insights from quantum mechanics, statistical mechanics, and condensed matter physics. By limiting each of them to the domain where they are used in practice and by combining them in a suitable manner, a coherent and plausible overall picture emerges that requires none of the various difficult-to-swallow interpretations of quantum mechanics. The discussions will be confined to nonrelativistic quantum mechanics, but the relevance for relativistic quantum theory is obvious and will briefly be mentioned at the end.

\section{Bound and unbound quantum systems}

The quantum mechanical description of bound and unbound quantum systems is fundamentally different, and this will become important in the discussion of transitions further below. A bound system, such as a molecule or a crystal, has a restricted range of relative motion and is fully specified by a Hamiltonian of the form (\ref{fullhamiltonian}), with eigenfunctions given by the equation
\begin{equation} \hat H |n\rangle = E_n |n\rangle\, . \end{equation} 
Since a bound system has a finite extension, the spectrum of energy eigenvalues is discrete. Its time evolution can therefore be written as
\begin{equation}  |\psi(t)\rangle = \sum_n c_n e^{-iE_nt/\hbar}|n\rangle  \, ,\end{equation}
with the coefficients $c_n= \langle n|\psi(0)\rangle$ specifying the initial state.
Such bound systems (and only these, as will become more clear further below) are timeless objects, as the Hamiltonian summarizes all their properties, and as the time evolution is nothing but a rotation of a unit vector in a well-specified Hilbert space. Any change of such a system requires the coupling to an external environment and the inclusion of unbound particles (such as emitted photons) in the description.

In contrast, unbound quantum systems, such as free particles leaving an emitter or particles scattering on each other, require a continuum of modes for their description. Their time evolution is not well defined unless one specifies boundary conditions in space and time. The mathematical formalism that is employed in this case uses Greens functions or propagators. There are two types of propagators. The propagator that is used in the path integral formulation of quantum mechanics and can also be applied to bound systems is defined as
\begin{equation} K(\vec r,t;\vec r\,',t')=\langle\vec r\,|e^{-i\hat H(t-t')/\hbar}|\vec r\,'\rangle\,  . \end{equation}
The propagator that is used for the calculation of Feynman diagrams, for instance in scattering theory, is defined as
\begin{equation}  G( \vec r,t;\vec r\,',t') =\frac 1{i\hbar}\theta(t-t')K(\vec r,t;\vec r\,',t')\end{equation}
and satisfies the equation
\begin{equation}  
(i\hbar\partial_t-\hat H_{\vec r})G( \vec r,t;\vec r\,',t')=\delta(t-t')\delta(\vec r-\vec r\,')\, .\label{greens}
\end{equation}
The Hamilton operator $\hat H$ is usually that of a free particle, $\hat H_0$, but it can also include a classical potential that acts as an unchanging background to the time evolution. 
The propagator $G$ in equation (\ref{greens}) is a Greens function and breaks explicitely time reversal symmetry. It is used to express the time evolution of a particle in terms of localized source terms. Let us consider scattering theory as an example, with the scattering center being represented by a potential $V(\vec r)$. (This means that the original two-particle system has been reduced to a one-particle system by going to relative coordinates.)
By splitting the Hamiltonian in two terms, $\hat H = \hat H_0+V$, the Schr\"odinger equation of the particle can be written as
\begin{equation} (i\hbar\partial_t-\hat H_0)\,\psi = V\psi\,.  \end{equation}
Using the Greens function, one obtains for $\psi$ the implicit equation
\begin{equation}  \psi(\vec r,t) = \psi_0(\vec r,t) + \int G(\vec r,t;\vec r\,',t') V(\vec r\,',t')\psi(\vec r\,',t')\, d\vec r\,' dt' \, , \label{scattering}\end{equation}
which is usually solved perturbatively by expressing in the integral on the right-hand side the wave function $ \psi(\vec r\,',t')$ again by the initial free wave function plus an integral involving the Greens function, and truncating after the desired number of iterations. 

The underlying physical picture is the following: $\psi_0$ represents the incoming wave function (usually a wave packet) that knows nothing about the potential. It is determined by its causal past, which includes some preparation or emission process, plus possibly past interactions. The second term describes the effect of the scattering potential on the time evolution of the wave function, assuming that its effect starts to be felt at some initial time when the particle has come close enough to the scattering center and ends at some final time, after which the scattering process is complete. Note that this situation is fundamentally different from a closed system discussed above, where the potential $V$ that is contained in $\hat H$ is timeless and is necessary for specifying the eigenstates $|n\rangle$, which can be used as the basis of the Hilbert space in which the "rotation" that represents the time evolution is performed. 

To summarize so far, the quantum-mechanical treatment of bound systems is done by using a timeless representation of the system, while the treatment of unbound systems depends crucially on the concept of causality, since the wave function is expressed in terms of a state that is thought to be due to its causal past and by changes induced by the present causal factors, which are formalized as the (temporary) action of interaction potentials. The step function in the definition of the propagator implements this causality. In explicit expressions for the free propagator causality is implemented by an added term $i\epsilon$ in the denominator of the Fourier-transformed propagator, or by prescriptions in which direction the poles shall be circumvented when the Fourier transform from frequency and momentum space to position coordinates and time is performed.  The delta functions on the right-hand side of (\ref{greens}) furthermore implement the idea that all causal factors can be expressed as a superposition of point-like sources.  

All this means that time reversal symmetry is broken in unbound quantum systems. The time evolution of the wave function is fundamentally different in forward and backward time direction. There are no sources in the future, but the wave function becomes more and more spread in space as time increases. This situation is very similar to that in classical electrodynamics, where retarded potentials are calculated and all electromagnetic radiation is traced back to localized sources in the past\cite{weinstein2011electromagnetism}.

All this also means that the time evolution in unbound systems cannot really be called "unitary", as it does not take place in a fixed, eternal Hilbert space, but interaction partners are introduced and taken away as time passes. If this is not done explicitely, it is done implicitely by applying the corresponding boundary conditions in space and time. The description by a fixed Hamiltonian and a formally unitary time evolution is used as a good approximation for a limited time interval only. 

There is in fact a third class of systems that is a hybrid between these two classes, namely finite-temperature macroscopic systems. Let us take a microcanonical system, which is isolated from the rest of the world. It has a finite size and thus a discrete spectrum. It is confined by some external potential, as for instance a gas in a box. If the system is a crystal, the relevant degrees of freedom are phonons, which are again confined in a "box", which is limited by the boundaries of the crystal. On the other hand, due to the considerable energy content and the large size of the system, the spectrum is very close to being continuous, and textbooks always assume that the energy uncertainty in the system is so large that is covers a huge number of energy eigenstates. In this sense we are dealing with an (almost) unbound system that one might rather describe by the concepts used in scattering theory. The time evolution of particles would then be described by some initial state that is due to the causal past of the particle and the interaction with nearby particles that affect its further time evolution over a limited time interval. As mentioned above, the two views are fundamentally incompatible, and we will discuss finite-temperature systems more carefully further below. As we will see, they hold the key for understanding the quantum to classical transition. 

In the following, we will first discuss the quantum-mechanical descriptions of bound systems ("structures"), then state transitions in bound systems and the coupling to a continuum, and then finite-temperature systems. Bringing the insights gained from these systems together, we will be able to see a coherent overall picture.

\section{Structures}

A structure is a bound many-particle system, such as an atom, a molecule, a crystal, a metal, a glass. It is a world on its own for many quantum-mechanical calculations.
Since a bound system is finite, it has a discrete energy spectrum. A good estimator of the density of energy eigenstates is Bohr's quantization rule which states that there is one eigenstate per phase space volume $h^{3N}$. For integrable systems that are expressed in terms of action-angle variables, Bohr's rule means the action variable of different energy eigenstates differs by integer multiples of Planck's constant $h$. 

The simplest structure is the hydrogen atom. It is a fully integrable two-body system. When isolated from the rest of the world, its energy is conserved. However, due to the interaction with their environment, hydrogen atoms make transitions between eigenstates by emitting or absorbing photons. These transitions will be discussed further below. The description of the hydrogen atom as an isolated system is thus only an approximation. The natural state for a hydrogen atom surrounded by vacuum is the ground state. 

Problems involving more than two particles cannot be exactly solved. This means that there is no fully analytical treatment based on all nuclei and electrons for any other many-particle system. One has to rely on approximative methods such as perturbation theory and variational calculus, for instance Hartree-Fock calculations and density functional theory. These methods calculate the ground state (or excited states) of the electrons, presupposing the potential created by the nuclei. This is the so-called Born-Oppenheimer approximation which exploits the fact that nuclei are much heavier than electrons.
However, as is pointed out by several authors\cite{primas1998emergence,chibbaro2014reductionism}, the Born-Oppenheimer approximation introduces classical features into quantum mechanics as it presupposes well localized nuclei. A fully quantum mechanical calculation of the ground state of a molecule cannot yield localized nuclei, as we shall explain in the next section. The assumption that nuclei are localized is inspired from the observation that molecules have well-defined shapes, which in turn is a feature that is due to a finite-temperature environment. We will discuss the mechanism by which temperature localizes atoms further below.  

Since molecules and solids cannot be treated exactly, they are treated using effective Hamiltonians that are capable of describing the low-lying excitations of structures, such as vibrations and rotations. In crystals (and also in disordered solids and liquids) these excitations are phonons. These Hamiltonians include usually harmonic oscillator potentials which are obtained by assuming small deviations of the atoms from their equilibrium positions. Just as in the Born-Oppenheimer approximation, a structure consisting of well localized atoms is presumed. The cause of the harmonic potential  around the equilibrium positions is the attractive interaction mediated by electrons. Low-lying electronic excitations, such as in molecules are metals, are also described by defining appropriate effective Hamiltonians that contain the relevant degrees of freedom. 

It becomes thus clear that the Hilbert spaces that are used for quantum mechanical calculations in structures are never the full Hilbert spaces of the original $N$-particle system. This is emphasized in particular by solid state physicists\cite{anderson1972more,leggett1992nature,laughlin2000theory}. They are based on a small subset of all possible degrees of freedom. The other degrees of freedom are not included in the calculation of the time evolution or are taken into account as external, classical potentials. 

In structures where energy levels become pretty dense (i.e., in macroscopic objects), excitations can be localized in a small part of the system and are then called "quasiparticles". These quasiparticles are formally treated exactly like free elementary particles in the vacuum\cite{falkenburg2015quasi}. Their propagation and interaction is described using Feynman diagrams and propagators that rely on the concept of causality. This shows again that finite-temperature macroscopic systems combine features of bound and unbound systems and need a carefull discussion (see section \ref{thermal}).  

\section{External potentials}
Many quantum mechanical calculations involve an external potential $V(\vec r)$ in the Schr\"odinger equation. This is very useful in particular for teaching quantum mechanics, as it allows the discussion of examples that are solvable. The two-body problem with a distance-dependent potential between the two particles can be reduced to a one-body problem with an external potential, but systems with more particles and mutual interactions cannot be reduced to a problem with an external potential\cite{ellis2012limits}. 

Nevertheless, external potentials are taken as a useful approximation for real physical situations that involve more particles, for instance atoms in quantum wells or the atoms of a gas in a box. The quantum well or the box are structures that consist of many atoms, which together produce the confining potential.  These potentials are usually treated as classical potentials, for instance electrostatic potentials. On a microscopic level, classical electrostatic potentials have charge densities as source terms. If the quantum well or the box were described in terms of the wave functions of their electrons and nuclei, these charge densities would be proportional to the absolute value squared of these wave functions, $\varrho(\vec r)= q \psi(\vec r)^*\psi(\vec r)$. Similarly, static magnetic fields are due to current densities, which are in quantum mechnics expressed as $\vec j(\vec r)=\frac{q\hbar}{2mi}\left[\psi(\vec r)^*\vec\nabla\psi(\vec r)-\psi(\vec r)\vec\nabla\psi(\vec r)^*\right]$. Other potentials, such as the van-der-Waals potential can also be expressed by quantum mechanical expectation values that contain the products $\psi(\vec r)^*\psi(\vec r)$ This means that the wave function of the atoms that constitute the confining structure enter the Schr\"odinger equation of the confined particle(s) in a nonlinear way. Furthermore, the confined particle does not act back on the confining structure. This situation, which has proven to describe observations well, is very different from what one would obtain if one did set up the full Schr\"odinger equation of the structure and the confined particles, which must be linear in all one-particle wave functions (if a product basis is chosen). This fully microscopic quantum mechanical treatment would thus lead to an entanglement of the confined particles with the atoms of the external structure, and to a destruction of phase coherence of the confined particles. 

Further below, we will give criteria under what circumstances the environment of a quantum system is appropriately described by a classical external potential.  In order to make sense of external potentials, there must be a limit to the range of linear superposition and entanglement.

A similar situation arises when several macroscopic objects are considered. Their interaction is usually successfully described by classical potentials and forces. This means again that there must be limits to linear superposition and entanglement so that each structure can be considered as a separate whole. 

\section{The formation of a structure is not a unitary process}

The modelling of structure formation requires two things: the first is an environment that confines the building blocks of the structure such that they come close together and can interact. Examples are the formation of a crystal in an undercooled liquid, the synthesis of biomolecules in biological cells, or the formation of hydrogen atoms in a sufficiently dense early universe. Since the structure shall be a world by itself that is independent of the rest of the world (to a good approximation), the quantum-mechanical description of structure formation cannot involve a combined unitary time evolution of all particles of the structure and the environment.  Rather, the confining environment must have the properties of a classical potential. The second thing that is needed for the description of structure formation is the coupling to a continuum that can take up the energy that is given away as the parts go into the bound state. Since the interactions considered in this paper are of electromagnetic nature, the continuum is in many cases the electromagnetic vacuum, and the emitted energy packet is a photon. If the environment is a material medium (e.g. a crystal or a liquid), phonons or other quasiparticles can replace the photon. Additionally, also atoms or molecules can be emitted during structure formation, for instance when biological macromolecules are synthesized in biological cells. Emission processes will be treated later, and it will be argued that their description is not unitary. After the structure has been formed, both these ingredients are taken away, as only the internal processes within the structure become part of the Hamiltonian that is used to describe the structure. The confining environment and the coupling to a continuum do not form part of the Hilbert space and the unitary time evolution of the structure. 

As an example, let us consider the formation of a crystal in more detail. The environment in which the crystal is formed is a liquid in which the atoms or molecules diffuse without being confined to a specific location, with the macroscopic (classical!) variables density and temperature setting the conditions for crystal formation. When an atom (or molecule) is added to the growing crystal structure, it makes a transition from diffusion in a medium to a bound state, where it becomes part of the vibrational degrees of freedom of the crystal. Furthermore, the energy difference is given to the environment as a photon or phonon. Let as put together the tools that are available for a quantum mechanical modelling of this process:  A particle in the liquid is described as a localized wave packet built by superposition of a continuum of modes. The most detailed existing description is given by quantum mechanical molecular dynamics simulations, which use a quantum description only locally, and otherwise classical interactions\cite{marx2000ab} (see Section 7 below for a brief description of these simulations). The emission of the photon or phonon is described by Fermi's golden rule or QFT, coupled to the transition of the particle from the liquid to the solid, which would probably be modelled as a two-level system or a tunneling process. The final degrees of freedom in the formed crystal are described by lattice vibrations. This shows that the transition from the liquid to the solid is far from being described by a unitary time evolution in an overarching many-particle Hilbert space. Instead, quantum mechanical descriptions capture only a small part of the system, and the used Hilbert spaces are completely different when describing the initial situation, the transition, and the final situation. Furthermore, as discussed also in the next section, the quantum mechanical treatment of transitions involves explicitely a breaking of time-reversal symmetry and a nonunitary projection on the final state when the transition probability (or rate) is calculated.         

Let us return to the general consideration of the formation of a structure. Quantum mechanics has a conceptual problem at explaining how structures, which break symmetries by assigning specific (relative) positions to atoms, can exist at all. Very generally, it is not possible that the confining potential has a symmetry that the ground state of the structure does not have.  
 The quantum mechanically correct ground state would then be not the structure as we know it, i.e. with localized atoms, but a superposition of all equivalent orientations of the structure in the environment. For instance, if the structure is a H$_2$O molecule (which has an angle of approx. 108 degrees between the two connections from the O atom to the H atoms) and the confinement has a cuboid symmetry, then there are (at least) 2 configurations of the molecule in the confinement
 that have exactly the same energy and that are related by a reflection on one of the three symmetry planes. In classical physics, each of these symmetry-broken state would be a ground state, because the atoms of the molecule must be localized. In quantum mechanics, this cannot be the exact ground state, because time evolution would result in a tunneling between the two same-energy states. But if a temporal change occurs, a state cannot be an eigenfunction of the Hamiltionan.   
 By forming symmetric and antisymmetric linear superpositions of these degenerate states $\psi_1$ and $\psi_2$, 
  \begin{equation}   
\psi_S=\frac 1{\sqrt 2}(\psi_1+\psi_2)\, , \qquad
\psi_A=\frac 1{\sqrt 2}(\psi_1-\psi_2)\, ,   
  \end{equation}  
 one obtains new states that have a different energy due to the electrostatic interactions between the superimposed wave functions. Both expectation values
 $$ 
\langle \psi_S|\hat H|\psi_S\rangle = \langle \psi_1|\hat H|\psi_1\rangle + \Re \langle \psi_1|\hat H|\psi_2\rangle $$
and
$$
\langle \psi_A|\hat H|\psi_A\rangle = \langle \psi_1|\hat H|\psi_1\rangle - \Re\langle \psi_1|\hat H|\psi_2\rangle
 $$
are upper bounds to the ground state energy, and one of the two must be lower than $\langle \psi_1|\hat H|\psi_1\rangle$, which was the  ground state energy when atoms were required to be localized. We thus see that a symmetry-broken state cannot be the ground state. The ground state must have the symmetry of the environment. 

 If we omit the external potential and assume that the structure is placed in infinite space that is homogeneous and isotropic, exactly the same problem arises. The quantum-mechanical ground state of the structure must also be homogeneous and isotropic. For this reason localized wave packets that describe the center-of-mass motion of objects become broader and broader when they evolve according to the free-particle Schr\"odinger equation. Broken symmetries cannot be explained within quantum mechanics, but they require a classical environment (e.g. an external potential) that also breaks this symmetry. Since for large structures the energy difference between the symmetric and the localized state becomes very small, a very weak symmetry-breaking potential is sufficient \cite{van2015instability}. 
 Since the modelling of any structure that has more than one atom breaks symmetries by assigning to atoms specific relative positions, the existence of structures can thus not be explained by quantum mechanics alone.  
 
 As mentioned before, the Born-Oppenheimer approximation that is employed when calculating the configurations of molecules is based on classical concepts.   This also shows that the quantum-mechanical Hilbert spaces of structures depend on classical physics for their existence. 
 
\section{State transitions in structures}

Without coupling to an environment, a structure retains its energy, and if it is not in an eigenstate of the Hamiltonian, its time evolution is a rotation in Hilbert space. 

State transitions in structures are accompanied by the emission or absorption of a photon or some other (quasi)particle(s). The quantum mechanical description of such transitions involves two elements that break unitary time evolution. First, the Hilbert space of the structure is extended by adding environmental degrees of freedom. These have not been included before as the initial state is described using only the eigenstates of the structure. Second, the environmental degrees of freedom are removed again when a projection on the final state is done in order to calculate the transition probability or transition rate. 

The simplest way to describe a state transition within the framework of nonrelativistic quantum mechanics is Fermi's golden rule, which is based on first order time-dependent perturbation theory. Here, the external degrees of freedom are implemented indirectly by adding a time-dependent potential for a limited time. In the following, we will analyze the derivation of Fermi's golden rule in order to discuss explicitly how time-reversal symmetry and unitarity are broken in those calculations. 
More exact treatments\cite{fonda1978decay}, such as the approach by Wigner and Wei\ss kopf or full QED treatments, could be discussed in a similar way, with the same conclusion.  

Let $\hat H_0$ be the Hamiltonian of the structure (for instance a hydrogen atom), and $W(t)=W(e^{i\omega t}+e^{-i\omega t})$ the
interaction with the environment (for instance a coupling of the dipole moment of the atom to the field of an electromagnetic wave), which is assumed to act for a limited time. The following calculation describes for instance the transition of a hydrogen atom from an excited state into the ground state by the emission of a photon. The eigenstates
$\{ |n\rangle \} $ of the structure are given by $\hat H_0 |n\rangle = E_n |n\rangle$.  The system is supposed to be initially in one of the eigenstates, $|i\rangle$. The Hilbert space used for the description of the initial state is thus spanned by the eigenstates $\{|n\rangle\}$ of $H_0$, and past processes that let to the formation of the excited state and involved interactions with other particles are ignored. Establishing the initial state $|i\rangle$ is a nonunitary process, or a projection. 
Next, the interaction potential $W(t)$ is coupled into the system. In more complete treatments of photon emission, the photonic degrees of freedom would be included explicitely, thus extending the Hilbert space by those degrees of freedom. In our case, these additional degrees of freedom are considered indirectly by coupling to this classical potential, which adds an external world to the system. 

Under the action of the full Hamiltonian $\hat H = \hat H_0 + W(t)$, the initial state goes into a superposition
\begin{equation}
|\psi(t)\rangle = \sum_n c_n(t) e^{-iE_nt/\hbar}|n\rangle \, ,
 \end{equation}
where the time evolution of the coefficients $c_n(t)$ is given by
\begin{equation}  
\frac{\partial c_n(t)}{\partial t}= \frac 1 {i\hbar}\sum_k c_k(t) W_{nk}(t) e^{i\omega_{nk}t} \label{dcdt}
\end{equation}
with the matrix elements $W_{nk}(t)=\langle k|W(t)|n\rangle$
and the transition frequencies $\omega_{nk}=(E_k-E_n)/\hbar$.
At this stage, the calculation describes a unitary evolution that is reversible in time. 

Next, the following approximations are made: (i) For short times, we can set on the right-hand side $c_i=1$ and all other $c_k=0$. (ii) We focus on the time evolution of coefficients $c_f$ that belong to states $|f\rangle$ that satisfy $\omega \simeq \omega_{fi}$, i.e. that yield transition frequencies that are close to a resonance with the frequency of the interaction potential. This means that $|\omega - \omega_{fi}| \ll |\omega_{fi}|\, ,$ allowing the neglection of the term proportional to $e^{-i\omega t}$ in $W(t)$ (rotating wave approximation). With these two assumptions, we obtain
\begin{equation}
|c_f(t)|^2 \simeq \frac{\omega_{fi}^2}{\hbar^2} \left|\frac{\sin[(\omega_{fi}-\omega)t/2]}{(\omega_{fi}-\omega)/2}\right|^2\, .
\label{resonance}
\end{equation}
At this stage, we have still a time-reversible expression, even though the assumption that the initial state $|i\rangle$ does not get depleted is unrealistic and not valid for times that do not satisfy $t \ll \hbar/W_{fi}|$. 

Two steps that break time-reversal symmetry are made by the subsequent calculation, which involves an interpretation of $|c_{fi}|^2$ as a transition probability, and the continuum limit. By assuming that the final state is part of a continuum with a density of states $\varrho(E)$, over which an integration is performed, one obtains a transition probability at time $t$
\begin{equation} P(t)= \int |c_{fi}|^2(t) \varrho(E) dE  \, , \end{equation}
which by using the formula
\begin{equation}\int_{-\infty}^\infty \frac{\sin ^2 x}{x^2}dx = \pi   \end{equation}
gives the familar expression for the transition rate,
\begin{equation}   
W \equiv \frac{dP(t)}{dt} \simeq \frac{2\pi}{\hbar}W_{fi}^2\; \varrho(E_{f})\, . \label{fermi} \end{equation}
The projection $\langle f|$ onto the final state that is done when $|c_{fi}|^2$ is interpreted as a transition probability, is nonunitary: The interaction potential (or, in a QFT treatment, the electromagnetic modes) are taken away, so that the Hilbert space used to describe the system is again the one used initially, and within this Hilbert space only the state $|f\rangle$ is considered, assuming that an irreversible, stochastic collapse on one eigenstate of $H_0$ occurs.

The assumption that the final state is part of a continuum is valid for macroscopic structures or for unbound particles (as in scattering theory). When transitions in a smaller structure (for instance a hydrogen atom) are calculated, the final states of the structure are discrete, but the states for the emitted photon are part of a continuum. When this is taken into account explicitely (such as in the Wigner-Wei\ss kopf theory\cite{weisskopf1930berechnung}), a similar calculational step is performed by integrating over this continuum (which corresponds in our calculation to an integration over $\omega$ instead of $\omega_{fi}$), and the discussion made in the next paragraph applies to these calculations as well. 

Let us discuss the meaning of this last calculational step, which involves taking the continuum limit.   The right-hand side of equation (\ref{resonance}) diverges when the continuum limit is taken. The physical reason for this divergence is a resonance between the interaction potential and the transition frequency. In the absence of this resonance, i.e., as long as the energy spectrum is discrete, 
the time evolution (\ref{dcdt}) is unitary and can be continued to infinite times. Energy would in this case be transferred in a quasiperiodic manner between the different modes. This situation can be constructed for instance for an initially excited hydrogen atom by placing it in a cavity that has a discrete spectrum of electromagnetic modes. Initially, energy would be transferred from the atom to the electromagnetic modes, with those that are closer to the resonance receiving more energy. If time was long enough, the system would return periodically to the initial state. The larger the cavity, the denser the modes and the closer to the resonance, and the longer the recurrence time. This means that reversibility is present in this system when the limit $\omega \to \omega_{fi}$ is taken after the limit $t \to \infty$ is taken. However, the last step that leads to (\ref{fermi}) reverses the order in which these limits are taken as the continuum limit is taken at finite time. This changes fundamentally the physics of the system. Instead of hovering forever among a limited number of states, the energy packet now leaves the initial state for good. 

The difference between these two ways of taking the limits is similar to the difference between the two types of time evolution described above in bound versus unbound systems.  The reason why propagators of unbound particles need the added $i\epsilon$ is a resonance. The mathematical meaning of this procedure is again an exchange of the continuum limit and the large-time limit, and the physical interpretation is again the transition from a reversible time evolution in a well-defined limited Hilbert space to an irreversible time evolution where a quantum state is determined by its sources in the past. The fact that resonances and a continuous spectrum introduce irreversibility has been emphasized for a long time by the Prigogine school \cite{petrosky1993poincare,petrosky1997liouville}.

The reverse process, the absorption of a photon by a structure, is not described by a unitary time evolution either. Nor is it the time-reversed situation of the emission process. The wave function of the emitted photon has an angular distribution that depends on the type of transition. The reversed process would be an incoming wave function with exactly this angular distribution, which would become fully absorbed by the structure. (In classical electrodynamics, such a process would be described using advanced potentials). The photon would thus be described by a "source" that lies in the future instead of the past. Instead, photons that become absorbed in structures have been emitted from another structure and are described by a wave function that has this emission process as its source. When it interacts with the other structure, it gets entangled with it. A definite transition of the structure cannot be described by this unitary time evolution, but requires again  "measurement" or "collapse" event. Calculations of absorption processes, which employ QFT, implement this insight by performing a projection on the desired final state, which is again a nonunitary process. 



\section{The important role of temperature} \label{thermal}

We have already mentioned above that finite-temperature macroscopic systems share features with bound systems as well as with unbound systems. Furthermore, they emit black-body radiation, which means that there are state transitions in thermal systems that emit and absorb photons. Let us consider the photon emission process in more detail. If we consider the system in a way that is analogous to the hydrogen atom, we would expect that photon emission and absorption is accompanied by transitions between energy eigenstates of the system. However, this idea becomes incoherent when the rate of state transitions is large, because this means a short lifetime and a large energy uncertainty of the levels, so that the levels overlap and cannot be considered anymore as distinct. Also, photon emission from macroscopic systems is generally assumed to happen locally and not globally. Otherwise the spectrum and the radiative power per unit area would not be independent of the system size. All this means that the idea that a finite-temperature many-particle system behaves like a coherent quantum mechanical whole with a global unitary time evolution and global coupling to the external world cannot be upheld without running into contradictions. The alternative picture derived from unbound systems that the atoms (or phonons, or any other relevant excitations) are exposed to a temporal sequence of limited-time interactions seems more appropriate. Accordingly, condensed matter physics uses the same formalism as quantum field theory, with propagators and perturbation calculations that implement causality. However, as mentioned in the discussion of Fermi's golden rule, the scattering of atoms in a gas or phonons in a crystal is not completely described unless the projection on a final state is performed. Otherwise the next scattering event could not be described in the same way, with a pretty well localized particle in the initial state. Furthermore, if the projection on a localized final state was not performed, the state of the many-particle system would again evolve to a coherent, global whole, which we have already ruled out. All this means that we have to conclude that the dynamics of a finite-temperature many-particle system involves localization or "collapse" events, which limit the length scale over which the description by a wave function is appropriate. These "collapse" events introduce a stochastic component into the description of the system, which ties in nicely with the stochastic concepts applied in statistical mechanics that convey the idea that energy is randomly exchanged between the different degrees of freedom, leading to a maximum-entropy state. Additional arguments for  continuous localization of atoms in a thermalized gas can be found in \cite{tenreasons,howmanybits}.

The length scale over which a wave function is coherent and capable of linear superposition is obtained from statistical mechanics\cite{kittel1980thermal}. It is given by the thermal wavelength, which is $\lambda_{th}= h/\sqrt{2\pi m k_B T}$ for atoms of mass $m$ in a gas, and $\lambda_{th}=(ch)/(2\pi^{1/3}k_BT) $ for phonons with the sound velocity $c$. There are two intuitive ways to derive their order of magnitude by simple arguments: the first consists in simply taking the thermal de Broglie wavelength $\lambda$, which is obtained from the relations $E \sim k_BT$  and $E=E(p)$ according to the energy-momentum relation that is valid for the considered particle, with $p=\hbar k = h/\lambda$.  The second consists in calculating the size of a cell in phase space that allows transitions with a change in energy of the order of $k_BT$. For a gas of atoms for instance, the density of states within a spatial cell of volume $l^3$ is determined from $\varrho(E)dE=l^3 \, 3p^2dp/\hbar^3$, due to the rule that there shall be one state in a phase space volume $\hbar^3$ (if spin is neglected). Using $E=p^2/2m$ and setting $E \sim k_BT$ and $dE \sim k_BT$ and $p=h/\lambda$, one obtains the desired expression. For massless particles, the equivalent calculation is done using $E=pc$. 

We have thus come to the conclusion that finite-temperature systems continuously localize their atoms and quasiparticles and that they are described by wave functions only locally and approximately. This solves many puzzles raised by a purely quantum mechanical description. In particular, the interaction with atoms that are much further away than the thermal wavelength must be described by a classical potential. As we have seen, this is necessary requirement for the formation of structures and for the existence of well limited and localized objects in space that interact via classical potentials. Below, we will show that this also solves the measurement problem.

Finally, let us discuss the system sizes and energies required for the breakdown of the unitary description of quantum mechanics. There is certainly a broad range of parameters over which the system is neither purely quantum mechanical nor purely classical. The requirements for the description of the thermalized macroscopic system outlined in this section are the negligibility of the influence of the walls or boundaries (which lead to a quasi continuous energy spectrum) and the coupling to a continuum of photonic degrees of freedom (which produces the black body spectrum). This means that the mean free path for the relevant excitations must be much smaller than the system size and that the system must be intransparent for photons of the wave length that correspond to the temperature. This points to system sizes of the order of micrometers to millimeters if the temperature is of the order of the room temperature.

\section{Quantum events and quantum measurement} 

The framework established so far enables us to understand the nature of quantum "events", which are nonunitary and stochastic changes in the way particles are arranged or move. Their theoretical description is fundamentally different from the timeless unitary evolution of bound quantum systems within a well-defined Hilbert space, as they involve changes of the Hilbert space itself.  

Examples for events (or cascades of events) are the following:  An electron is knocked out of a conductor by the photoelectric effect, or by the impact of another electron in a photomultiplier; a photographic plate becomes black at some spot where it is hit by an electron; a DNA molecule suffers a double strand break as a ionized particle track goes through it; atoms move by activated jumps over barriers in a supercooled liquid; atoms become permanently rearranged in a normal liquid; an atom becomes incorporated in a growing crystal and takes a fixed position; hydrogen atoms undergo nuclear fusion to helium in the sun. 

Events happen all the time in finite-temperature systems. Let us consider a simple liquid or gas. Beyond distances of the order of the thermal wavelength, the surroundings of an atom are perceived via classical potentials. Since these potentials change with time, this creates an ever changing environment for the atoms. This means that the Hilbert space that is appropriate to describe the time evolution of the wave function, also changes all the time, as local state transitions occur that are accompanied by the transfer of energy to other regions in the system, possibly combined with the emission or absorption of photons. Each of these state transitions, where energy is transferred to distances beyond the thermal wavelength, is an event.  

Events also happen in structures that are coupled to such a thermal environment. Let us take as an example a hydrogen atom in the excited state, which emits a photon and goes to the ground state. As long as the hydrogen atom (or the photon) does not interact with its environment, it remains entangled with the emitted photon, and the direction of the recoil of the atom is not specified. However, the thermal environment localizes the atom, thus fixing its direction of recoil and destroying the entanglement with the photon.  

Events also happen when particles enter a thermal environment from outside and deposit energy in it, either becoming integrated and thermalized, or leaving the system again with less energy. Let us use this type of event in order to illustrate the difference between the view suggested in this paper and standard decoherence theory. According to decoherence theory, unitary time evolution takes the combined system, which consists of the incoming particle in state $|i\rangle$ and the finite-temperature environment $|\mathcal{E}_0\rangle$, from the initial product state $|i\rangle |\mathcal{E}_0\rangle$ to a later-time state
\begin{equation}
|\psi_f\rangle = \sum_f c_f |f\rangle |\mathcal{E}_f\rangle \, ,\label{entangle}
\end{equation}
which is an entangled state of the possible particle states $|f\rangle$ with the corresponding environment states. The natural, or preferred, basis for the states $|f\rangle$ is those of localized particles since the interaction potential between the particle and the environment is position dependent. If the environmental degrees of freedom are ignored, the state of the particle is then described by a reduced density matrix $\varrho_{red }$ that is obtained by taking the trace over the environmental degrees of freedom. Assuming that the environmental states that cooccur with the different states $|f\rangle$ are mutually orthogonal, it is given by
\begin{equation}
\varrho_{red}\equiv \sum_{\mathcal{E}}\langle \mathcal{E}|\psi_f\rangle\langle \psi_f|\mathcal E \rangle = \sum_f |c_f|^2 |f\rangle\langle f| \, .\label{rhored}
\end{equation}
Although decoherence theory provides many useful insights, it cannot explain the measurement process, as it gives a (incoherent, classical) superposition of the possible measurement outcomes, and not only one outcome\cite{adler2003decoherence,schlosshauer2005decoherence}.
In contrast, the arguments presented in the previous section suggest that the final state $|\psi_f\rangle$ cannot involve an entangled environment, as all its particles are localized. The final state of the environment must therefore take the form of a product state of the different small regions of size $\lambda_{th}$. This means that the sum (\ref{entangle}) can in fact only contain one term, and that the final state of the incoming particle is described by only one state $|f\rangle$, and not by the entire sum (\ref{rhored}). Furthermore, the description by a wave function $|f\rangle$ is only approximately correct, as dynamics is nonunitary. In this respect, the present approach is related to continuous collapse theories\cite{bassi2013}, which however do not model explicitely a thermal environment but assume implicitely the presence of additional degrees of freedom that justify the inclusion of stochastic and dissipative terms in the time evolution of the wave function. 

Quantum measurements are a particular class of quantum events (or cascades of events). They require a sufficiently long-lived change in structure that is strongly correlated with the measured quantity and that is observable on a macroscopic scale, for instance by a pointer etc.\cite{ellis2012limits}. Therefore, an amplification mechanism is involved. 
Quantum measurements always involve the transient or permanent localization of the measured particle within a small spatial region, for instance on a photographic plate or along the track in a cloud chamber.  The place or time of this localization event gives the desired information about the quantity of interest (for instance the spin in a Stern-Gerlach experiment). Furthermore, this localization event changes the state and/or arrangement of the atoms in the measurement device. As we have seen, such a change cannot occur without coupling to a thermal environment, and indeed all measurement devices are macroscopic objects at finite temperature.

\section{Conclusions}

The arguments presented in this paper have shown that the theoretical description used for quantum mechanical systems presupposes a classical environment in various ways: The ignoring of interactions far in the past or future, the assigning of specific positions to the atoms of molecules and solids, the simplified use of external potentials instead of the full quantum mechanical interaction with all external atoms, and pre- and postselection of simple initial and final states would not be possible in a world of universal unitary time evolution. The dependence of quantum mechanics on classical concepts and environments is also emphasized by other authors\cite{primas1998emergence,weinstein2001absolute,ellis2012limits,chibbaro2014reductionism}.

In order to bridge the gap between the quantum and classical descriptions, we have argued that macroscopic finite-temperature systems have an intrinsic length scale beyond which they are classical. This means that particles or quasiparticles are confined within this scale and have no entanglement and no quantum coherence beyond this distance and that the interactions due to atoms outside this distance can be captured by classical potentials. Consequently, particles coming in contact with thermalized systems become also localized and lose the entanglement with the source that produced or emitted them. In contrast to decoherence theory, which holds that entanglement with the environment exists but is invisible when only subsystems are observed, this paper argued that this entanglement is not present at all beyond the thermal wavelength, as there are limits to the description of many-particle systems by unitary time evolution and by wave functions.

It is important to note that this length scale only applies to those degrees of freedom that participate in the random energy exchange that generates the finite temperature. Long-range quantum states, as in high-$T_C$ superconductors or entanglend photon pairs in quantum experiments, are not ruled out as long as their coupling to the environmental degrees of freedom is weak enough. An interesting open question is the range and lifetime of quantum correlations in biological systems\cite{panitchayangkoon2010long}, in which many degrees of freedom are far from thermodynamic equilibrium. 

Quantum mechanics, as it is employed in practice, changes Hilbert spaces by adding, removing, and changing degrees of freedom as structures are formed or changed or destroyed, and as particles are emitted or absorbed from structures. These changes of structures, which require the changes of Hilbert spaces, are stochastic events, which involve dissipation whenever the energy difference is given irreversibly to the environment, for instance by emitting a photon. This means that quantum mechanics is strongly influenced in a top-down way by the larger-scale changes occurring on the level of structures and described to a large extent by classical physics. Ultimately, all those changes are due to the fact that the universe is far from thermodynamic equilibrium. As emphasized by George Ellis\cite{ellis2012limits}, the cosmological arrow of time is the source for the other arrows of time that occur on smaller scales.   

The arguments presented in this paper propose that the quantum field theoretical calculations done in condensed matter physics require a finite-temperature environment that sets a cutoff scale for quantum correlations and justifies the assumption of localized particles. Since the formalism of QFT in relativistic quantum physics is extremely similar, this suggests that a cutoff scale for quantum superposition exists also in the quantum vacuum. The fact that large-scale interactions are described by a classical potential (gravity) also points to such a cutoff. However, so far there is not yet a generally accepted theory that combines gravity with quantum physics. Nevertheless, there are various indications that the concept of temperature could be important at solving this problem\cite{bekenstein1973black,hawking1974black,jacobson1995thermodynamics}.  
\acknowledgements{I thank Linda Reichl for alerting  me to the irreversibility of transitions to continua 20 years ago, and  Andrzej Banburski, Linquin Chen, George Ellis, Matteo Smerlak, Lee Smolin, and Steve Weinstein for helpful discussions and comments while preparing this manuscript. This work was supported in part by Perimeter Institute of Theoretical Physics. Perimeter Institute is supported by the Government of Canada  through the Department of Innovation, Science and Economic Development and by the Province of Ontario through the Ministry of Research, Innovation and Science.}

\end{document}